\documentclass[epj]{svjour}
\usepackage{tipa}
\usepackage{amssymb}
\usepackage{multirow,setspace,times,amssymb,amsmath,graphicx,color,rotating,subfigure,url}
\usepackage{lineno}
\usepackage[square,sort&compress,comma]{natbib}

\begin{document}

\title{Preferred numbers and the distribution of
trade sizes and trading volumes in the Chinese stock market}

\author{Guo-Hua Mu\inst{1,2,3} \and Wei Chen\inst{4} \and J{\'a}nos Kert{\'e}sz\inst{3,5,}\thanks{\emph{e-mail}: kertesz@phy.bme.hu} %
\and Wei-Xing Zhou\inst{1,2,6,7,8,}\thanks{\emph{e-mail}:wxzhou@ecust.edu.cn} %
 }

\institute{School of Business, East China University of Science and
Technology, Shanghai 200237, China \and School of Science, East
China University of Science and Technology, Shanghai 200237, China
\and Department of Theoretical Physics, Budapest University of
Technology and Economics, Budapest, Hungary \and Shenzhen Stock
Exchange, 5045 Shennan East Road, Shenzhen 518010, China \and
Laboratory of Computational Engineering, Helsinki University of
Technology, Espoo, Finland \and Research Center for Econophysics,
East China University of Science and Technology, Shanghai 200237,
China \and Engineering Research Center of Process Systems
Engineering (Ministry of Education), East China University of
Science and Technology, Shanghai 200237, China \and Research Center
on Fictitious Economics \& Data Science, Chinese Academy of
Sciences, Beijing 100080, China}
\date{Received: \today / Revised version: date}

\abstract{ The distribution of trade sizes and trading volumes are
investigated based on the limit order book data of 22 liquid Chinese
stocks listed on the Shenzhen Stock Exchange in the whole year 2003.
We observe that the size distribution of trades for individual
stocks exhibits jumps, which is caused by the number preference of
traders when placing orders. We analyze the applicability of the
``$q$-Gamma'' function for fitting the distribution by the
Cram\'{e}r-von Mises criterion. The empirical PDFs of trading
volumes at different timescales $\Delta{t}$ ranging from 1 min to
240 min can be well modeled. The applicability of the $q$-Gamma
functions for multiple trades is restricted to the transaction
numbers $\Delta{n}\leqslant8$. We find that all the PDFs have
power-law tails for large volumes. Using careful estimation of the
average tail exponents $\alpha$ of the distribution of trade sizes
and trading volumes, we get $\alpha>2$, well outside the L{\'e}vy
regime.
} 
\maketitle

\section{Introduction}
\label{S1:Intro}

A well-known adage says: It takes volume to move stock prices,
indicating that the trade size and trading volume contain much
information about the dynamics of price formation. The topic of
price-volume relationship has a long history in finance
\cite{Karpoff-1987-JFQA} and recently has been investigated at the
transaction level
\cite{Chan-Fong-2000-JFE,Lillo-Farmer-Mantegna-2003-Nature,Lim-Coggins-2005-QF,Naes-Skjeltorp-2006-JFinM,Zhou-2007-XXX}.
Furthermore, understanding the origin of power-law tails in returns
is an important issue. In the unified theory of Gabaix {\em{et al}}
\cite{Gabaix-Gopikrishnan-Plerou-Stanley-2003-Nature}, the power-law
tails of returns are related to the power-law tails of volumes,
where the institutional activity plays a crucial role
\cite{Gabaix-Gopikrishnan-Plerou-Stanley-2006-QJE,Gabaix-Gopikrishnan-Plerou-Stanley-2007-JEEA,Gabaix-Gopikrishnan-Plerou-Stanley-2008-JEDC}.
Zhou verified the relation at the transaction level, although the
values of exponents are remarkably different from those of Gabaix
{\em{et al}} \cite{Zhou-2007-XXX}. On the other hand, Farmer {\em{et
al}} have found that large price changes at the transaction level
are driven by liquidity fluctuations rather than by volume
\cite{Farmer-Gillemot-Lillo-Mike-Sen-2004-QF}. In particular, they
showed that the return distribution is closely related to the
distribution of gaps between the first two price levels on the order
book. Webber and Rosenow argued that large stock price fluctuations
can be explained quantitatively by taking into account both the
order flow and the liquidity \cite{Weber-Rosenow-2006-QF}. In
addition, Joulin {\em{et al}} also found that most price jumps are
induced by order flow fluctuations close to the point of vanishing
liquidity and volume plays a minor role
\cite{Joulin-Lefevre-Grunberg-Bouchaud-2008-XXX}. In any case, the
distributions of trading volume are of interest, also as part of
stylized facts.

The issue of the tail distribution of trade sizes is also
controversial. The probability density function of trade sizes has a
fat tail often described by a power law:
\begin{equation}
 f(v) \propto v^{-\alpha-1}~.
 \label{Eq:fv:PL}
\end{equation}
Gopikrishnan {\em{et al}} analyzed the transaction data for the
largest 1000 stocks traded on the three major US markets and found
that the distribution of trade sizes follows a power-law tail with
the exponent being $1.53\pm0.07$, and called this as the half-cubic
law
\cite{Gopikrishnan-Plerou-Gabaix-Stanley-2000-PRE,Plerou-Gopikrishnan-Gabaix-Amaral-Stanley-2001-QF,Plerou-Gopikrishnan-Gabaix-Stanley-2004-QF}.
The aggregated trading volumes at timescales from a few minutes to
several hundred minutes were found to have a power-law tail exponent
$1.7\pm0.1$
\cite{Gopikrishnan-Plerou-Gabaix-Stanley-2000-PRE,Plerou-Gopikrishnan-Gabaix-Amaral-Stanley-2001-QF,Plerou-Gopikrishnan-Gabaix-Stanley-2004-QF}
\footnote{It is noteworthy to point out that Farmer and Lillo
investigated three LSE stocks and found no clear evidence for
power-law tails \cite{Farmer-Lillo-2004-QF}.}. Maslov and Mills also
found that the trade sizes of several NASDAQ stocks have power-law
tails with the exponent close to 1.4 \cite{Maslov-Mills-2001-PA}.
Plerou and Stanley extended this analysis to other two markets (LSE
and Paris Bourse) and found quantitatively similar results across
the three distinct markets \cite{Plerou-Stanley-2007-PRE}. All these
tail exponents were found to be well below 2, within the L\'evy
regime.

Alternatively, Eisler and Kert{\'e}sz reported that the tail
exponents of the traded volumes at a timescale of 15 minutes for six
NYSE stocks are 2.2 and 2.8
\cite{Eisler-Kertesz-2006-EPJB,Eisler-Kertesz-2007-PA}. Racz, Eisler
and Kert{\'e}sz argued that the tail exponents of the trade size
were underestimated \cite{Racz-Eisler-Kertesz-2008-XXX}. They
investigated the 1000 most liquid stocks traded on the NYSE for the
same period as studied by Plerou and Stanley
\cite{Plerou-Stanley-2007-PRE} and found that the average tail
exponent is $2.02\pm0.45$. The tail exponents were found to be
outside the L\'evy regime for stocks in the Korean stocks
\cite{Lee-Lee-2007-PA} and the Chinese stocks \cite{Zhou-2007-XXX}.

The above mentioned studies concern with the tail behavior of the
trade size or trading volume distribution. There are also efforts
attempting to describe the whole distribution of trading volumes.
The normalized trading volumes of 10 top-volume NYSE stocks and
NASDAQ stocks at different timescales of 1 min, 2 min and 3 min were
fitted by the $q$-Gamma distribution
\cite{Tsallis-Anteneodo-Borland-Osorio-2003-PA}
\begin{equation}
 f_{qG}(v)= \frac{1}{Z} \left(\frac{v}{\theta}\right)^{\beta}
 \left[1-(1-q){\frac{v}{\theta}}\right]^{\frac{1}{1-q}}~,
 \label{Eq:qGamma}
\end{equation}
where
\begin{equation}
 Z=\int_{0}^{\infty}\left(\frac{v}{\theta}\right)^{\beta}
 \left[1-(1-q){\frac{v}{\theta}}\right]^{\frac{1}{1-q}}{\rm{d}}v
 \label{Eq:Z}
\end{equation}
is the normalization constant, $\theta$ and $\beta$ are positive
parameters, and the value of $q$ is larger than 1. The usage of the
$q$-Gamma distribution can be motivated from a stochastic dynamical
scenario
\cite{Queiros-2005-EPL,deSouza-Moyano-Queiros-2006-EPJB,Queiros-Moyano-deSouza-Tsallis-2007-EPJB}.
In the limiting case $q \rightarrow$ 1, the traditional Gamma
probability density function is recovered. The asymptotic behavior
of the $q$-Gamma distribution has a power-law form
(Eq.~(\ref{Eq:fv:PL})) with
\begin{equation}
 f_{qG}(v) \sim v^{1/(1-q)+\beta}.
 \label{Eq:f:qg:PL}
\end{equation}
Compared with Eq.~(\ref{Eq:fv:PL}), we find the asymptotic tail
exponent
\begin{equation}
\alpha^{'}={\frac{1}{q-1}}-\beta-1.
 \label{Eq:alpha:q:beta}
\end{equation}
Here, we use $\alpha^{'}$ to make the difference between the
empirical tail exponent and the fitting one. From the results of
reference \cite{Queiros-2005-EPL}, we find the tail exponent
$\alpha^{'}$ is greater than 2 for 10 high-volume NASDAQ stocks (for
which $\beta$ and $q$ are available) at different timescales of 1
min and 2 min.

In this paper, we focus on investigating the distributions of trade
sizes and trading volumes, based on limit order book data of 22
liquid stocks. The trade size $\omega\equiv\omega_{i}$ is defined as
the number of shares exchanged in trade $i$, and trading volume is
defined as the share volumes in a fixed time interval. Here we
consider two types of trading volumes based on different definitions
of the time interval, event time and clock time. The first one is
defined as the total volume traded in a fixed clock-time interval
$\Delta{t}$:
\begin{equation}
\Omega_{\Delta{t}} = \sum_{i=1}^{N_{\Delta{t}}}\omega_{i}~,
 \label{Eq:Qt}
\end{equation}
where $N_{\Delta{t}}$ is the number of transactions in a fixed
interval $\Delta{t}$. The second one is defined as the total volume
traded in a fixed event-time interval $\Delta{n}$:
\begin{equation}
\Omega_{\Delta{n}} = \sum_{i=1}^{\Delta{n}}\omega_{i}~,
 \label{Eq:Qn}
\end{equation}
where $\Delta{n}$ is the number of trades. The issue of trade sizes
is a special case when $\Delta{n}=1$. Our work deals with the
problem of describing the whole distributions and with the
estimation of the tail exponent.

The paper is organized as follows. After a brief description of the
data in Section \ref{S1:Data}, we investigate the distributions of
the trade sizes ($\omega$) and the two types of trading volumes
($\Omega_{\Delta t}$ and $\Omega_{\Delta n}$) in Sections
\ref{S1:TradeSize}, \ref{S1:TradingVolume1} and
\ref{S1:TradingVolume2}, respectively. Section \ref{S1:Conclusion}
concludes.

\section{Description of the data}
\label{S1:Data}

We used tick by tick data for 22 liquid stocks traded on the
Shenzhen Stock Exchange (SZSE) in the whole year 2003. The market
consists of three time periods on each trading day, namely, the
opening call action (9:15 AM to 9:25 AM), the cooling period (9:25
AM to 9:30 AM), and the continuous double auction (9:30 AM to 11:30
AM and 1:00 PM to 3:00 PM). In this paper, we consider only the
transactions occurring in the double continuous auction. The size of
each transaction is recorded in the data sets. In the Chinese stock
market, the size of a buy order is limited to a board lot of 100
shares or an integer multiple thereof, while a seller can place a
sell order with any size. The recorded trade size is in units of
shares.

The tickers of the 22 stocks investigated are the following: 000001
(Shenzhen Development Bank Co. Ltd: 887,741 trades), 000002 (China
Vanke Co. Ltd: 509,360 trades), 000009 (China Baoan Group Co. Ltd:
447,660 trades), 000012 (CSG holding Co. Ltd: 290,148 trades),
000016 (Konka Group Co. Ltd: 188,526 trades), 000021 (Shenzhen Kaifa
Technology Co. Ltd: 411,326 trades), 000024 (China Merchants
Property Development Co. Ltd: 133,586 trades), 000027 (Shenzhen
Energy Investment Co. Ltd: 313,057 trades), 000063 (ZTE Corporation,
265,450 trades), 000066 (Great Wall Technology Co. Ltd: 277,262
trades), 000088 (Shenzhen Yan Tian Port Holdings Co. Ltd: 97,195
trades), 000089 (Shenzhen Airport Co. Ltd: 189,117 trades), 000429
(Jiangxi Ganyue Expressway Co. Ltd: 117,424 trades), 000488
(Shandong Chenming Paper Group Co. Ltd: 120,097 trades), 000539
(Guangdong Electric Power Development Co. Ltd: 114,721 trades),
000541 (Foshan Electrical and Lighting Co. Ltd: 68,737 trades),
000550 (Jiangling Motors Co. Ltd: 346,176 trades), 000581 (Weifu
High-Technology Co. Ltd: 93,947 trades), 000625 (Chongqing Changan
Automobile Co. Ltd: 397,393 trades), 000709 (Tangshan Iron and Steel
Co. Ltd: 207,756 trades), 000720 (Shandong Luneng Taishan Cable Co.
Ltd: 132,233 trades), and 000778 (Xinxing Ductile Iron Pipes Co.
Ltd: 157,321 trades).

The 22 stocks investigated in this work cover a variety of industry
sectors such as financials, real estate, conglomerates, metals \&
nonmetals, electronics, utilities, IT, transportation,
petrochemicals, paper \& printing and manufacturing. Our sample
stocks were part of the 40 constituent stocks included in the
Shenshen Stock Exchange Component Index in 2003
\cite{Zhou-2007-XXX}.

Note that when we investigate trading volumes $\Omega_{\Delta t}$
and $\Omega_{\Delta n}$ of all 22 stocks, we have normalized the
trade sizes $\omega$ by the total number of outstanding shares to
account for share splits.

\begin{figure*}[htb]
\centering
\includegraphics[width=6.5cm]{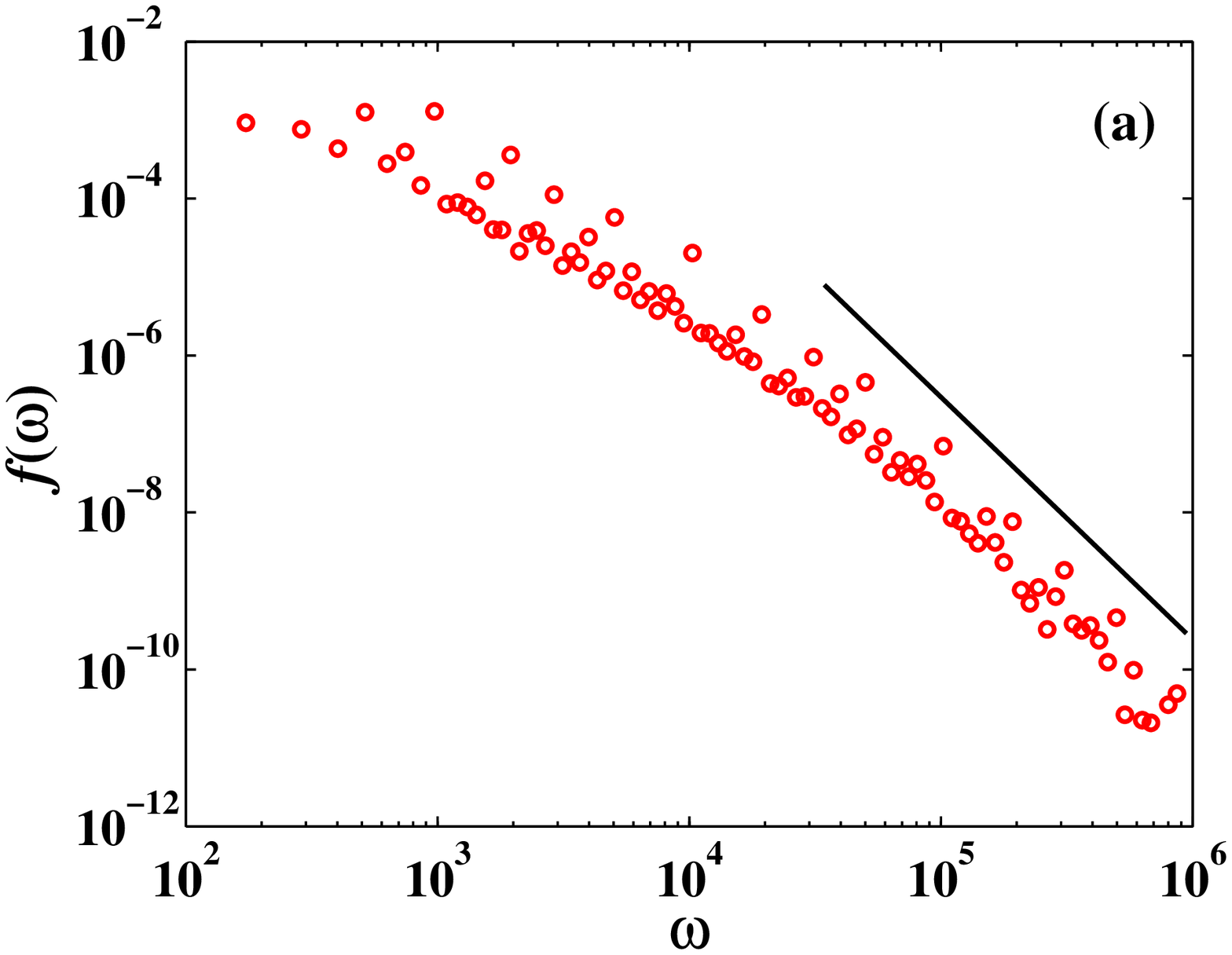}
\includegraphics[width=6.5cm]{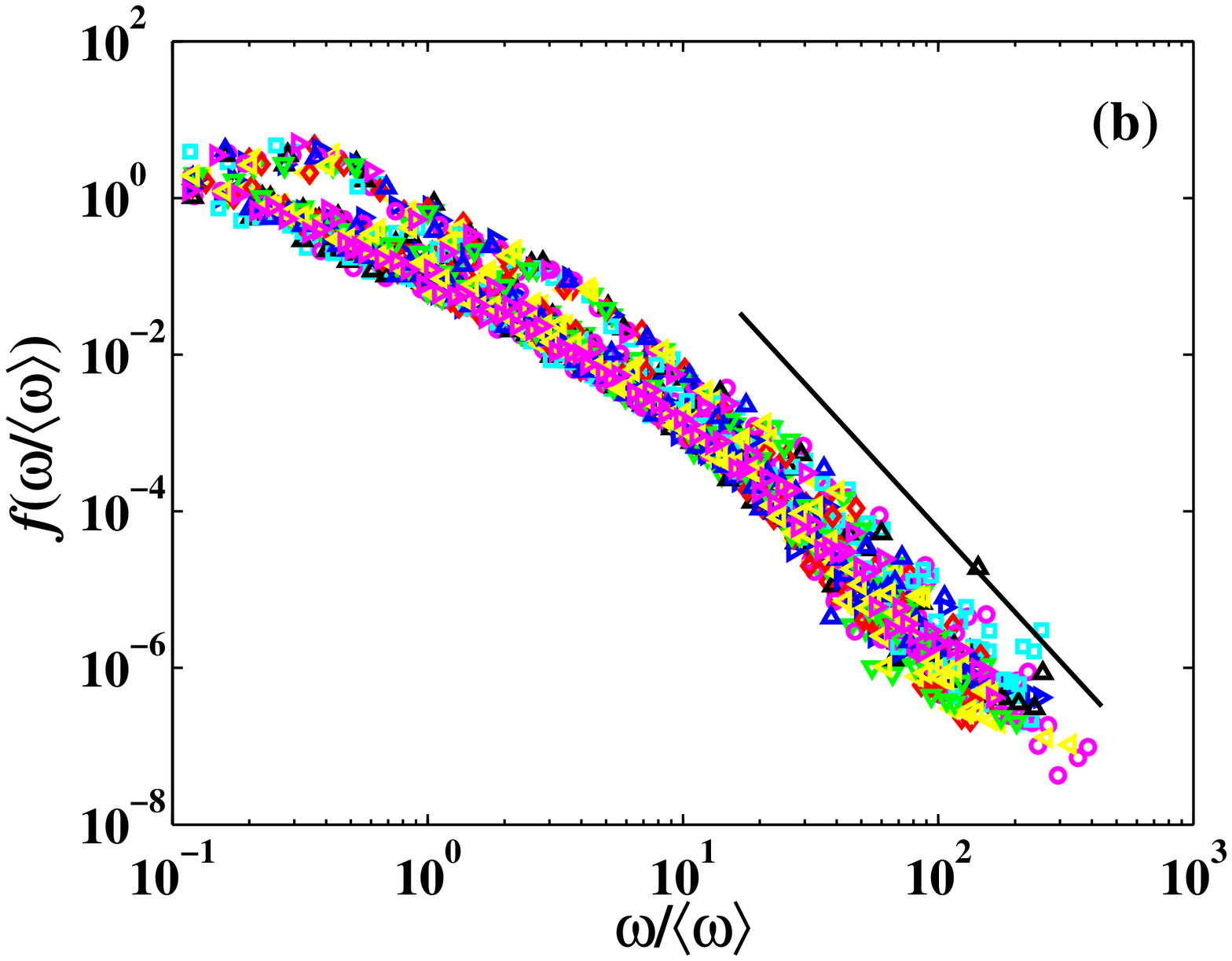}
\caption{\label{Fig:PDF:TradeSize} (Color online) (a) Empirical
probability density function $f(\omega)$ of the trade sizes of a
typical stock (code 000001). (b) Empirical probability density
distributions $f(\omega/\langle\omega\rangle)$ of the normalized
trade sizes of 22 liquid stocks. Here, we have normalized the trade
sizes $\omega$ by the total number of outstanding shares to account
for share splits, and $\langle\omega\rangle$ is the mean trade size
for individual stocks.}
\end{figure*}

\section{Trade size distribution}
\label{S1:TradeSize}

\subsection{Trade size distribution of individual stocks}

In this section, we study the trade size distribution of individual
stocks. Fig.~\ref{Fig:PDF:TradeSize}(a) displays the probability
density function (PDF) $f(\omega)$ of a typical stock (Shenzhen
Development Bank Co., LTD, code 000001) and
Fig.~\ref{Fig:PDF:TradeSize}(b) shows the PDFs
$f(\omega/\langle\omega\rangle)$ of the normalized trade sizes of 22
stocks, where $\langle\omega\rangle$ is the mean trade size for
individual stocks. The most intriguing feature is that the PDFs are
not continuous and there are evident jumps in the curves. The
outliers seem rather equidistant on the log scale. The reason is
that the traders seem to have a sense of {\em{number preference}},
which will be discussed in detail in Section \ref{S2:NumberPrefer}.
Therefore, the group of 22 PDF curves in
Fig.~\ref{Fig:PDF:TradeSize}(b) looks very thick. In the previous
work, the cumulative distribution of trade sizes is investigated. In
this way, the jumps are smoothened out. We note that this phenomenon
disappears for trading volumes.

\subsection{Number preference}
\label{S2:NumberPrefer}

\begin{figure}[htb]
\centering
\includegraphics[width=4.3cm]{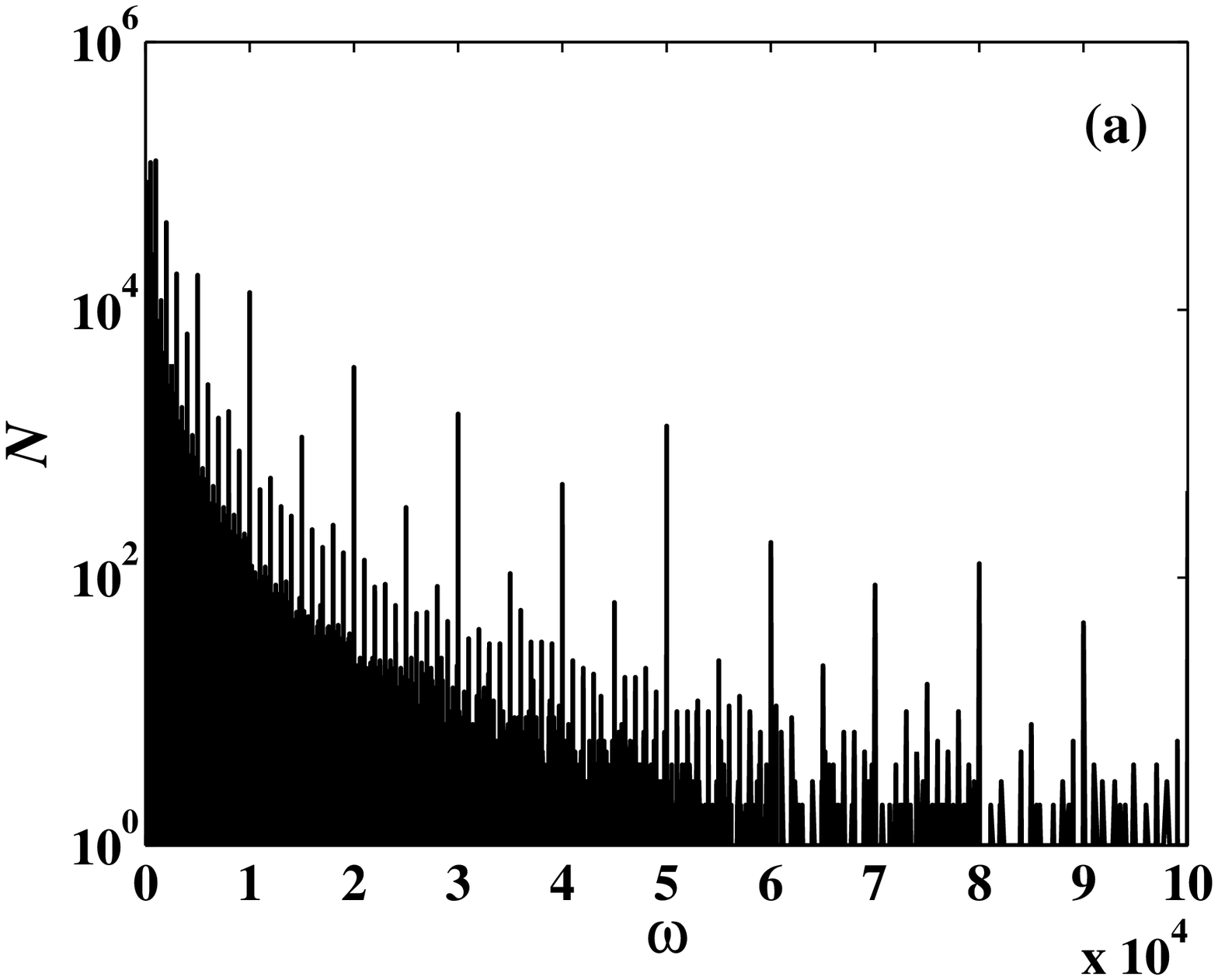}
\includegraphics[width=4.3cm]{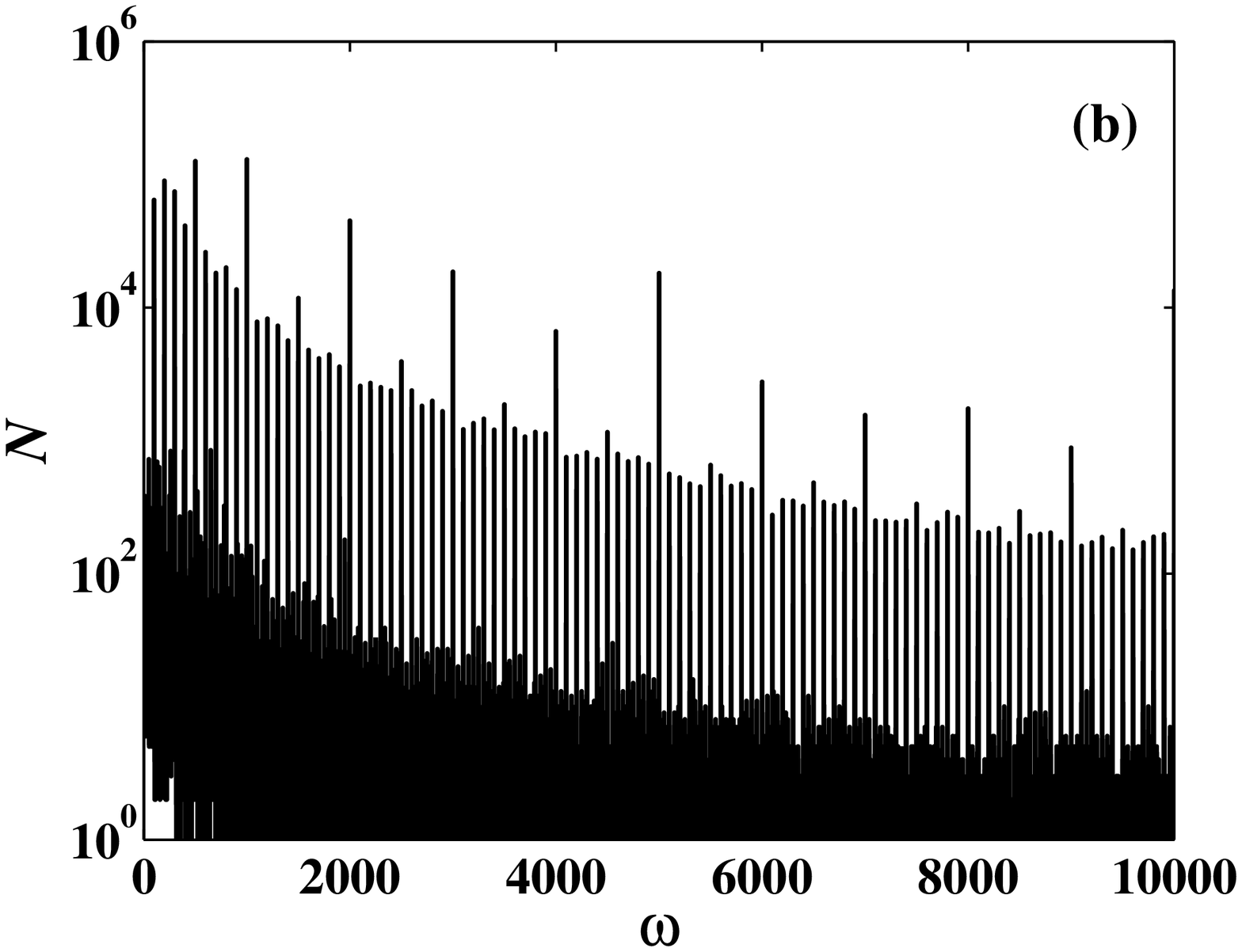}
\includegraphics[width=4.3cm]{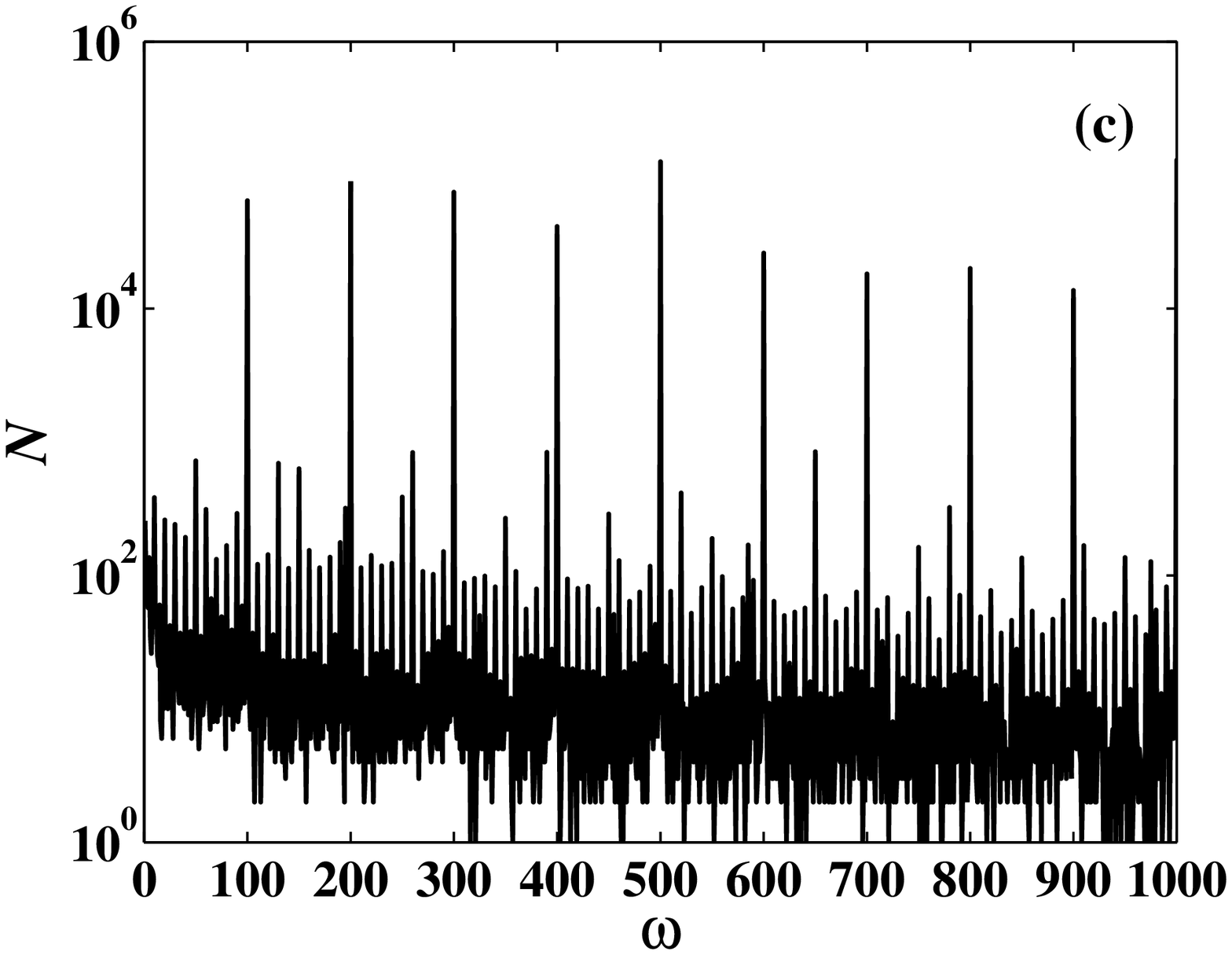}
\includegraphics[width=4.3cm]{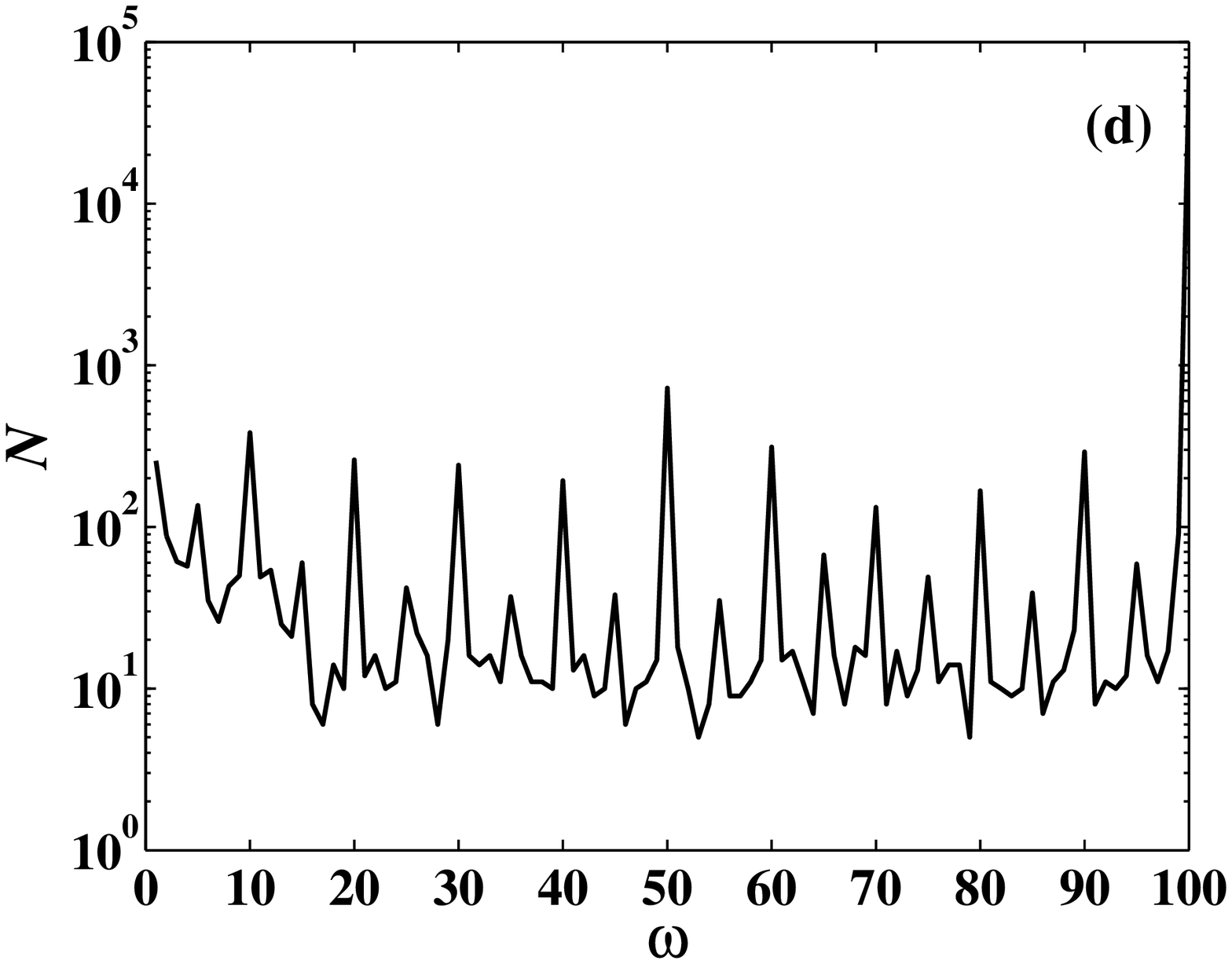}
\caption{\label{Fig:All_Nq} The number of transactions with the same
trade size as a function of the trade size for stock 000001.}
\end{figure}

Due to a variety of diverse natural and human factors, the numbers'
frequency of occurrence in human world shows an uneven behavior.
With the help of search engines, Dorogovtsev {\it et al} found the
occurrence frequencies of numbers in the World Wide Web pages are
very different, 777 and 1000, for example, occur much more
frequently than their neighbors
\cite{Dorogovtsev-Mendes-Oliveira-2006-PA}. This situation also
happens in the stock market. The order price placement shows
irrational preference of some numbers like 5, 10 or their multiples
\cite{Gu-Chen-Zhou-2008c-PA} in Chinese stock market. The first
observation can trace back to Niederhoffer
\cite{Niederhoffer-1965-OR} in 1965, showing the price are often on
integers, then on halves, on quarters, and Harris did more works on
it \cite{Harris-1990-JFQA}. Moreover, the frequency of first digit
in stock's price or return does not correspond to the frequency of
$\frac{1}{9}$ for each digit from 1 to
9~\cite{Pietronero-Tosatti-Tosatti-Vespignani-2000-PA,Giles-2007-AEL},
which also reflects people's number preference. Similar phenomenon
exists in the number of trades, as shown in Fig.~\ref{Fig:All_Nq}.
This is accordant to the result of Alexander and
Peterson~\cite{Alexander-Peterson-2007-JFE}. Fig.~\ref{Fig:All_Nq}
plots the number of transactions with the same trade size as a
function of the trade size for stock 000001 when the trade sizes are
among 1 to $10^5$ shares. It is observed that there are several
layers of spikes in these plots. Fig.~\ref{Fig:All_Nq}(a) shows the
first-layer spikes locating at $\omega=10^4k$ and the second-layer
spikes at $\omega=10^4(k+0.5)$, where $k=1,2,\cdots,9$. The
third-layer spikes at $\omega=10^3k$ and the fourth-layer spikes at
$\omega=10^3(k+0.5)$ can been seen in Fig.~\ref{Fig:All_Nq}(b).
Fig.~\ref{Fig:All_Nq}(c) and (d) depict more layers of spikes for
smaller trades. These spikes explain the jumps in the trade size
PDFs in Fig.~\ref{Fig:PDF:TradeSize}. Comparing
Fig.~\ref{Fig:All_Nq}(d) with other three plots, we find that the
magnitude of $N$ for $\omega<100$ is much lower than those for large
trades.

\subsection{Fitting the distribution}

We now fit the PDF of the normalized trade sizes
$v=\omega/\langle\omega\rangle$ of all the 22 stocks. The $q$-Gamma
distribution (Eq.~(\ref{Eq:qGamma})) is applied to model the PDF.
For comparison, we also adopt the $q$-exponential distribution
$f_{q}(v)$
\cite{Burr-1942-AMS,Tsallis-1988-JSP,Nadarajah-Kotz-2007-PA},
student distribution $f_{t}(v)$\cite{Blattberg-Gonedes-1974-JB} and
log-normal distribution $f_{l}(v)$, which are defined as follows:
\begin{equation}
f_{q}(v)=
\frac{1}{\theta}\left[1-(1-q){\frac{v}{\theta}}\right]^{\frac{q}{1-q}}~,
 \label{Eq:qExp}
\end{equation}
where $\theta>0$ and $q>1$,
\begin{equation}
f_{t}(v)=
\frac{n^{n/2}}{B(\frac{1}{2},\frac{n}{2})}\left[n+H(v-x)^2\right]^{-(n+1)/2}\sqrt{H}~,
 \label{Eq:t}
\end{equation}
where $B(\cdot, \cdot)$ is the ``beta function'', location parameter
$x$, scale parameter $H$ and degrees of freedom parameter $n$ are
positive, and
\begin{equation}
f_{l}(v)=
\frac{1}{v\sigma\sqrt{2\pi}}\exp\left[{\frac{-(\ln{v}-\mu)^2}{2\sigma^2}}\right]~,
 \label{Eq:lognormal}
\end{equation}
where $\mu$ and $\sigma$ are corresponding mean and standard
deviation respectively. The $q$-exponential (Eq.~(\ref{Eq:qExp}))
has an asymptotic power-law tail whose tail exponent is
\begin{equation}
 \alpha^{'} = {\frac{q}{q-1}}-1~.
 \label{Eq:alpha:q}
\end{equation}

\begin{figure}[htb]
\centering
\includegraphics[width=8cm]{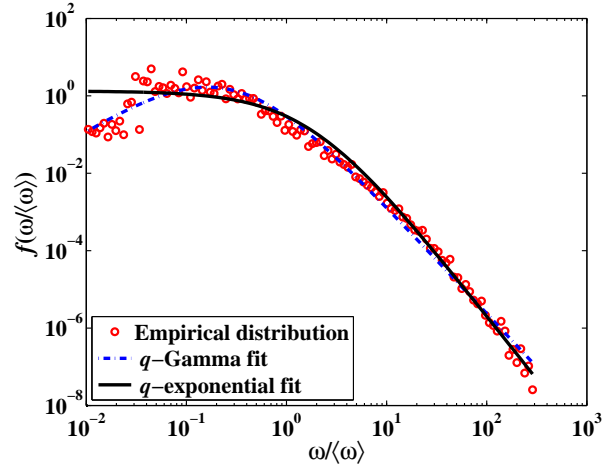}
\caption{\label{Fig:PDF:TradeSize:fit} (Color online) Fitting the
empirical PDF $f(\omega/\langle\omega\rangle)$ of trade sizes of all
the 22 stocks, which have already been normalized by the total
number of outstanding shares. The dashed curve is the $q$-Gamma fit
using Eq.~(\ref{Eq:qGamma}) with $\theta=0.07$, $\beta=1.52$,
$q=1.22$ and $\chi=0.57$ and the solid curve is the $q$-exponential
fit using Eq.~(\ref{Eq:qExp}) with $\theta=0.7$, $q=1.45$ and
$\chi=0.74$. The corresponding tail exponents $\alpha^{'}$ are 2.03
and 2.22 respectively.}
\end{figure}

The empirical PDF and two fits ($f_{qG}$ and $f_q$) are illustrated
in Fig.~\ref{Fig:PDF:TradeSize:fit}, because student and log-normal
fits deviate considerably from the empirical PDF, especially the
student function. It is visible that $q$-exponential gives bad fit
for small values, but a better fit than $q$-Gamma for large values.
Using taboo search and least-squares estimator, we obtain that the
parameters of the $q$-Gamma are $\theta=0.07$, $\beta=1.52$,
$q=1.22$ and $\chi=0.57$, while that of the $q$-exponential are
$\theta=0.7$, $q=1.45$ and $\chi=0.74$. According to
Eq.~(\ref{Eq:alpha:q:beta}) and Eq.~(\ref{Eq:alpha:q}), the tail
exponents $\alpha^{'}$ in the two models are 2.03 and 2.22,
respectively.

It is clear from Fig.~\ref{Fig:PDF:TradeSize:fit} that the $q$-Gamma
distribution fits remarkably well the empirical PDF for the whole
interval, while the $q$-exponential does not. Here, the
Cram\'{e}r-von Mises criterion is used for judging the
goodness-of-fit of the probability distribution compared to a given
distribution~\cite{Darling-1957-AMS}, which is given by
\begin{equation}
C_{M}^{2}=n\int_{-\infty}^{+\infty}\left[F(v)-F^{*}(v)\right]^{2}{\rm{d}}F(v)~,
 \label{Eq:CvM1}
\end{equation}
where $F^{*}$ is the empirical cumulative distribution function, and
$F$ is the corresponding theoretical distribution. In one-sample
applications, the function can be described as
follows~\cite{Pearson-Stephens-1962-Bm,Stephens-1970-JRSSB},
\begin{equation}
C_{M}^{2}=\frac{1}{12n}+\sum_{i=1}^{n}\left[\frac{2i-1}{2n}-F(v_{i})\right]^{2}~,
 \label{Eq:CvM2}
\end{equation}
where $n$ is the sample size. At the significance level 0.01, we
find that $C_{M}^{2}$ for $q$-Gamma is smaller than the critical
value using Eq.~\ref{Eq:CvM2}, while the $q$-exponential's is
larger. So we can accept the hypothesis that the trade sizes data
$\omega$ come from $q$-Gamma distribution rather than
$q$-exponential distribution.

\subsection{Determination of the tail exponent}

Calibrating the $q$-Gamma distribution already provides an estimate
of the tail exponent. However, a careful comparison of the fitted
curve with the empirical curve in the tail of
Fig.~\ref{Fig:PDF:TradeSize:fit} indicates that this estimate might
be biased. Here, we utilize six methods to estimate the tail
exponents, including the least-squares estimator (LSE) based on a
linear fit of a power law, Hill's estimator (HE) that is a
conditional maximum likelihood estimator \cite{Hill-1975-AS},
Meerschaert and Scheffer's estimator (MSE) based on the behavior of
moments \cite{Meerschaert-Schffer-1998-JSPI}, Clauset, Shalizi and
Newman's estimator (CSNE) based on maximum likelihood methods and
the Kolmogorov-Smirnov statistic
\cite{Clauset-Shalizi-Newman-2007-XXX}, Fraga Alves's estimator
(FAE) that is a location invariant Hill-type estimator
\cite{Alves-2001-EX}, and shift-optimized Hill estimator by Racz and
Kertesz (RKE) that can handle data
shifts~\cite{Racz-Kertesz-2008-xxxx}, which is an extension of the
CSNE. The resultant estimates of the tail exponent using LSE, HE,
MSE, CSNE, FAE and RKE are $2.33\pm0.08$, $1.54\pm0.07$,
$1.75\pm0.003$, $2.23\pm0.23$, $2.59\pm0.39$ and $2.13\pm0.23$,
respectively. As pointed out by Racz {\em{et al.}}, the MSE is
unable to predict exponent above 2, and systematically
underestimates the tail exponent
\cite{Racz-Eisler-Kertesz-2008-XXX}, while the FAE results in a high
variance \cite{Racz-Kertesz-2008-xxxx}.

\section{Trading volume at different timescales}
\label{S1:TradingVolume1}

\subsection{Bulk distribution}

\begin{figure*}[htb]
\centering
\includegraphics[width=6.5cm]{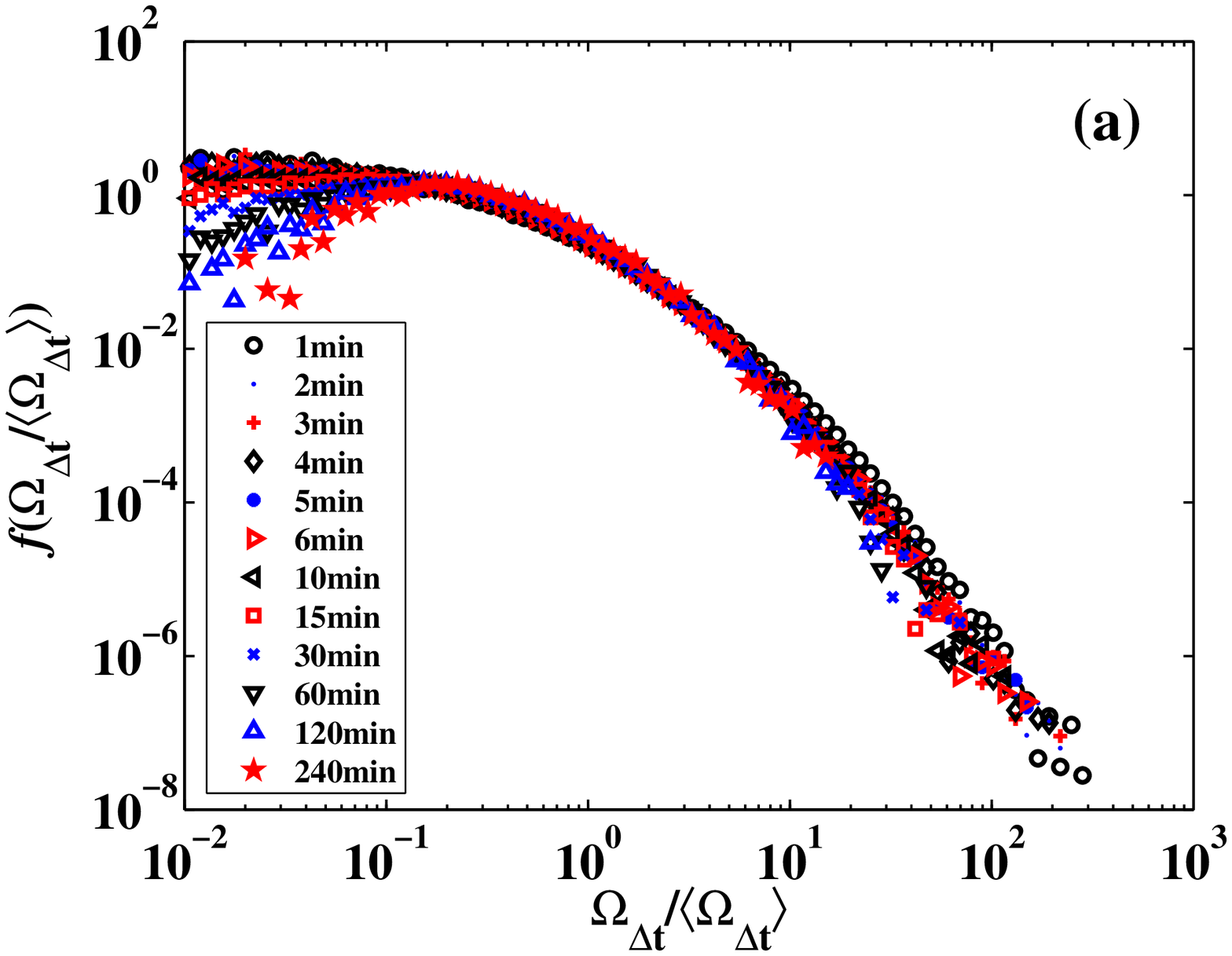}
\includegraphics[width=6.5cm]{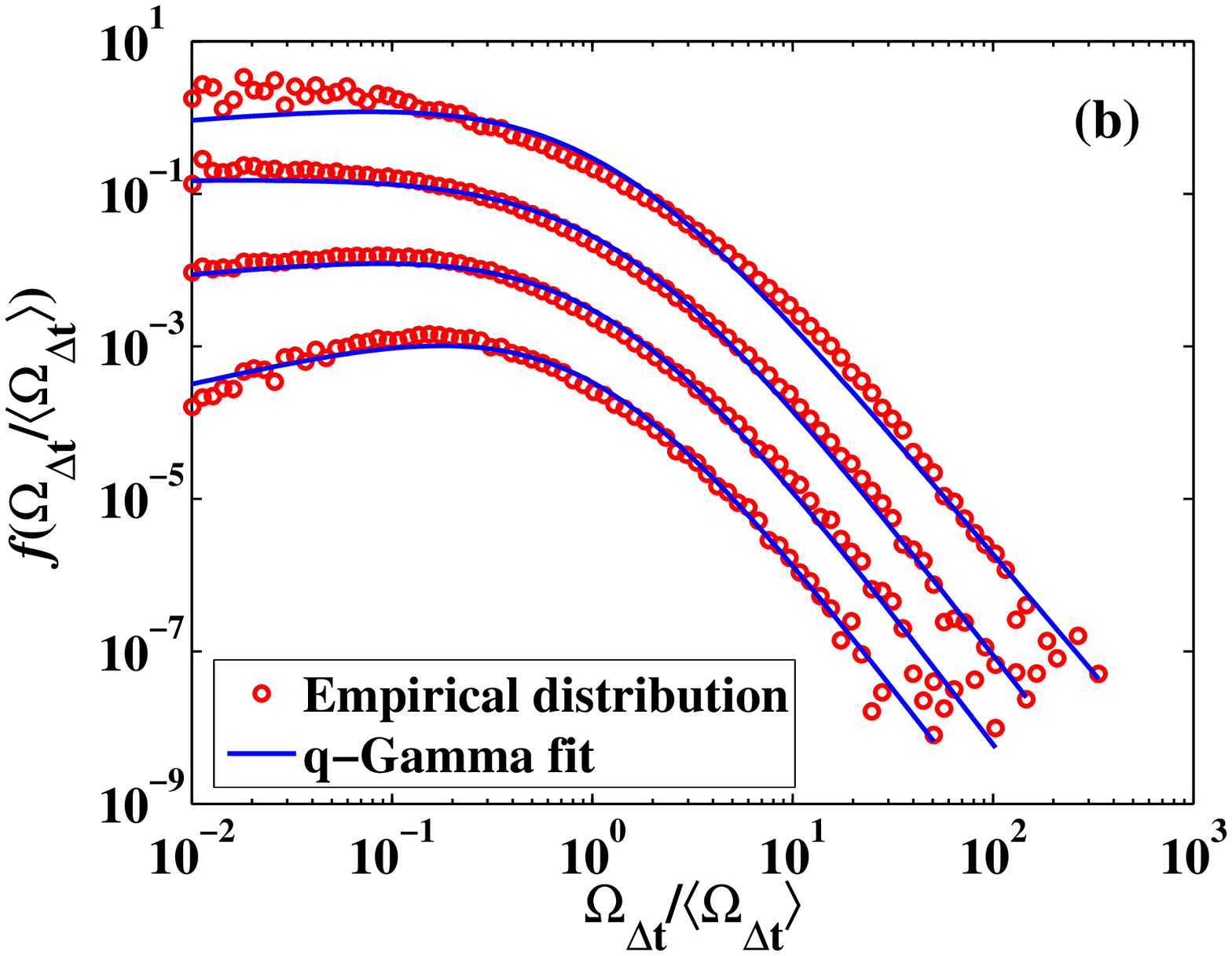}
\caption{\label{Fig:Qt} (Color online) (a) Probability density
functions of the normalized trading volumes at different timescales
$\Delta t$ ranging from 1 min to 240 min (1 trading day). (b)
Fitting the PDFs using $q$-Gamma (solid curves). The timescales from
top to bottom are $\Delta t=$ 1 min, 5 min, 15 min and 60 min,
respectively. We have shifted the curves for clarity.}
\end{figure*}

The distributions of trading volumes $\Omega_{\Delta t}$ in
different clock-time intervals $\Delta t$ have been studied for
several other stock markets. The PDFs of the normalized trading
volumes at different timescales are depicted in Fig.~\ref{Fig:Qt}
for the Chinese stocks. The timescale $\Delta t$ ranges from 1 min
to 240 min (1 trading day). The width of the PDF decreases with the
increase of timescale. In addition, there is no scaling in these
PDFs. For small values of $\Omega_{\Delta{t}}$, the density
decreases with $\Delta{t}$. For large values of
$\Omega_{\Delta{t}}$, the PDFs decay in power-law forms.

We also fitted these PDFs using $q$-Gamma, $q$-exponential and
log-normal densities except the student density.  Only the results
of $q$-Gamma fit for four typical cases with $\Delta{t}=$ 1 min, 5
min, 15 min and 60 min are illustrated in Fig.~\ref{Fig:Qt}(b). It
is clear that the $q$-Gamma functions fit remarkably the empirical
PDFs, which is also reported for other markets with $\Delta{t}=$
1min, 2 min and 3 min
\cite{Tsallis-Anteneodo-Borland-Osorio-2003-PA,Queiros-2005-EPL,deSouza-Moyano-Queiros-2006-EPJB,Queiros-Moyano-deSouza-Tsallis-2007-EPJB}.
We use the Cram\'{e}r-von Mises criterion again and find the values
of $C_{M}^{2}$ for $q$-Gamma function are smaller than the critical
value at the significance level 0.01, when $\Delta t$ ranges from 1
min to 240 min.

\begin{table}[htp]
 \caption{\label{Tb:FittingPDF:Qt} Calibration of the $q$-Gamma model to trading volumes at different clock
 timescales.}
 \medskip
 \centering
 \begin{tabular}{ccccccc}
 \hline \hline
  $\Delta{t}$&& $\theta$ & $\beta$ &  $q$ & $\chi$  &  $\alpha^{'}$ \\\hline%
    1 min   &&    0.41   &  0.15  &  1.29  &  0.52  &  2.30         \\%
    2 min   &&    0.32   &  0.22  &  1.28  &  0.52  &  2.35         \\%
    3 min   &&    0.42   &  0.01  &  1.29  &  0.42  &  2.44         \\%
    4 min   &&    0.37   &  0.08  &  1.28  &  0.44  &  2.49         \\%
    5 min   &&    0.41   &  0.04  &  1.28  &  0.44  &  2.53         \\%
    6 min   &&    0.40   &  0.06  &  1.28  &  0.42  &  2.51         \\%
    10 min   &&   0.39   &  0.11  &  1.27  &  0.21  &  2.59         \\%
    15 min   &&   0.31   &  0.26  &  1.26  &  0.29  &  2.59         \\%
    30 min   &&   0.19   &  0.57  &  1.26  &  0.27  &  2.28         \\%
    60 min   &&   0.12   &  1.09  &  1.23  &  0.30  &  2.26         \\%
    120 min   &&  0.09   &  1.51  &  1.22  &  0.27  &  2.04         \\%
    240 min   &&  0.07   &  2.05  &  1.20  &  0.31  &  1.95         \\%
    \hline \hline
 \end{tabular}
\end{table}

In contrast, the $q$-exponential function and log-normal function
have significant deviations from the empirical distributions, so we
do not display them in Fig.~\ref{Fig:Qt}(b). But $q$-exponential
grasps well the tail behaviors for small $\Delta t$ and log-normal
grasps well for large values for large $\Delta t$. This is not
surprising since the $q$-exponential (Eq.(\ref{Eq:qExp})) is a
monotonically decreasing function and close to power law for large
values while empirical PDFs show symmetrically for large values for
large $\Delta t$, tending to log-normal distribution. However, we
cannot distinguish the $q$-Gamma and $q$-exponential functions in
the tails. The estimated parameters using nonlinear least-squares
regression for all the PDFs are listed in Table
\ref{Tb:FittingPDF:Qt}. All but one tail exponents $\alpha^{'}$ are
larger than two.

\subsection{Tail exponents}

In order to further confirm that the tail exponents of trading
volumes at different timescales are consistently outside the
L{\'e}vy regime, we also adopt the six different estimators used in
Section \ref{S1:TradeSize} for the tail exponents. The results are
presented in Table~\ref{Tb:FittingPDF:Qt_tail}. For comparison, we
also show the $\alpha^{'}$ values obtained from the $q$-Gamma
fitting. For all the six tail exponent estimators, the value of
$\alpha$ trends up with the increase of $\Delta{t}$. All the tail
exponents obtained based on the MSE are less than two, while CSNE,
LSE, FAE and RKE give tail exponents larger than two. For the Hill
estimator, $\alpha$ is less than two when $\Delta{t}\leqslant6$ and
greater than two when $\Delta{t}\geqslant10$. Again, we argue that
the tail exponents of the trading volumes at different timescales do
not belong to the L{\'e}vy regime.

\begin{table*}[htp]
 \caption{\label{Tb:FittingPDF:Qt_tail} Estimating tail exponents of trading volumes at different clock timescales.}
 \medskip
 \centering
 \begin{tabular}{ccccccccccc}
 \hline \hline
  $\Delta{t}$&& $\alpha^{'}$  &&   HE           &   MSE          &  CSNE           &  LSE          &        FAE        &           RKE\\%
    \hline
    1 min   &&      2.30      && $1.71\pm0.03$  & $1.75\pm0.006$ &  $2.10\pm0.24$  & $2.43\pm0.09$ &    $2.80\pm0.27$  & $2.17\pm0.07$ \\%
    2 min   &&      2.35      && $1.83\pm0.07$  & $1.76\pm0.007$ &  $2.11\pm0.26$  & $2.52\pm0.06$ &    $2.93\pm0.16$  & $2.34\pm0.10$ \\%
    3 min   &&      2.44      && $1.88\pm0.10$  & $1.77\pm0.014$ &  $2.30\pm0.42$  & $2.50\pm0.08$ &    $2.93\pm0.14$  & $2.21\pm0.08$ \\%
    4 min   &&      2.49      && $1.91\pm0.07$  & $1.76\pm0.011$ &  $2.61\pm0.25$  & $2.58\pm0.11$ &    $2.97\pm0.25$  & $2.41\pm0.11$ \\%
    5 min   &&      2.53      && $1.94\pm0.09$  & $1.78\pm0.016$ &  $2.62\pm0.24$  & $2.56\pm0.11$ &    $3.04\pm0.19$  & $2.20\pm0.11$  \\%
    6 min   &&      2.51      && $1.95\pm0.10$  & $1.78\pm0.009$ &  $2.60\pm0.34$  & $2.71\pm0.11$ &    $3.01\pm0.24$  & $2.49\pm0.09$ \\%
    10 min   &&     2.59      && $2.20\pm0.10$  & $1.79\pm0.016$ &  $2.67\pm0.36$  & $2.81\pm0.21$ &    $3.13\pm0.35$  & $2.55\pm0.10$  \\%
    15 min   &&     2.59      && $2.28\pm0.11$  & $1.79\pm0.027$ &  $2.80\pm0.49$  & $2.58\pm0.31$ &    $3.37\pm0.36$  & $2.76\pm0.19$  \\%
    30 min   &&     2.28      && $2.46\pm0.22$  & $1.81\pm0.06$ &  $3.12\pm0.52$  & $2.78\pm0.25$ &    $3.17\pm0.33$   & $2.66\pm0.21$   \\%
    60 min   &&     2.26      && $2.45\pm0.28$  & $1.82\pm0.018$ &  $3.54\pm0.62$  & $2.40\pm0.49$ &    $3.50\pm0.55$  & $2.81\pm0.18$   \\%
    120 min   &&    2.04      && $2.46\pm0.29$  & $1.83\pm0.021$ &  $3.35\pm0.53$  & $2.47\pm0.37$ &    $3.26\pm0.50$  & $2.86\pm0.20$   \\%
    240 min   &&    1.95      && $2.51\pm0.24$  & $1.85\pm0.061$ &  $4.99\pm0.71$  & $3.57\pm0.14$ &    $3.08\pm0.64$  & $3.00\pm0.50$    \\%
    \hline \hline
 \end{tabular}
\end{table*}

\section{Trading volume distribution for multiple trades}
\label{S1:TradingVolume2}

\begin{figure*}[htb]
\centering
\includegraphics[width=6.5cm]{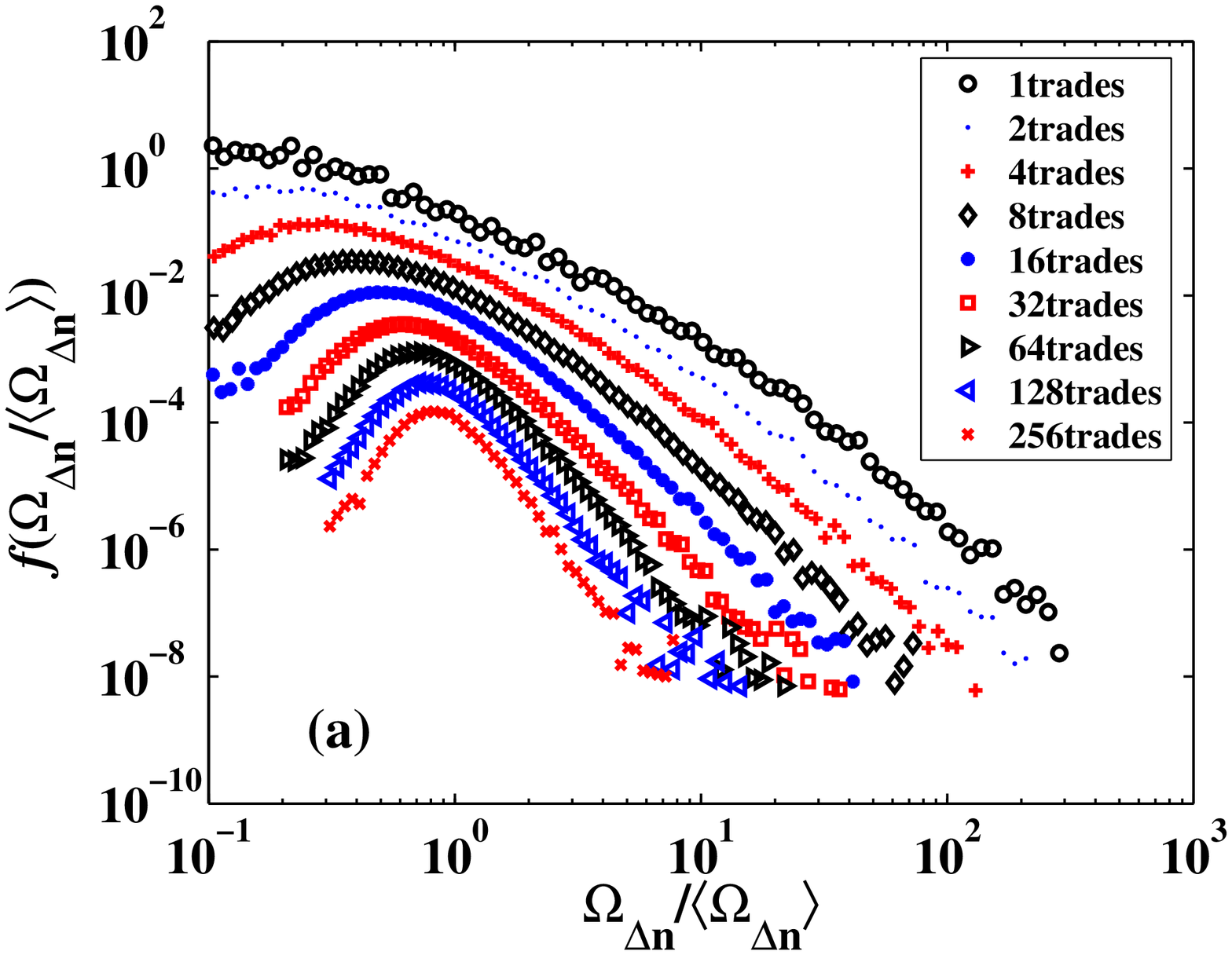}
\includegraphics[width=6.5cm]{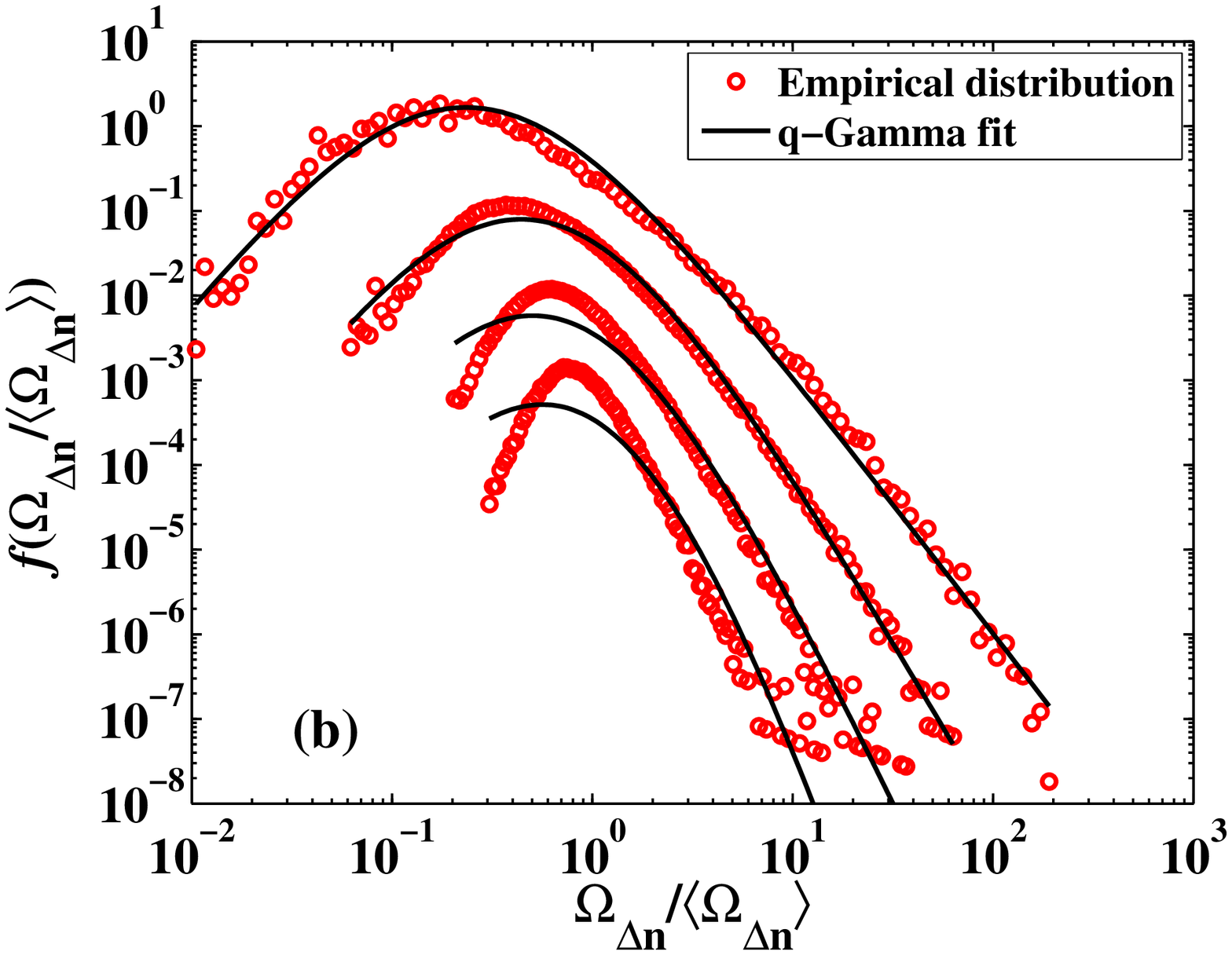}
\caption{\label{Fig:Qn} (Color online) (a) Probability density
functions of the normalized trading volumes at different timescales
$\Delta n = 2^k$ with $k=0,1,2,\cdots,8$. (b) Fitting the PDFs using
$q$-Gamma (solid curves). The timescales from top to bottom are
$\Delta n=$ 2, 8, 32, 128, respectively. We have shifted the curves
for clarity.}
\end{figure*}

We now study the distributions of trading volumes $\Omega_{\Delta
n}$ for multiple transactions. The trading volumes are normalized by
the averages for different $\Delta{n}$. Fig.~\ref{Fig:Qn}(a)
illustrates the empirical PDFs for $\Delta n = 2^k$ with
$k=0,1,2,\cdots,8$. It is found that the PDFs have power-law tails.
We use $q$-Gamma functions to fit the PDFs, since the other
distribution has significant deviations from the empirical
distribution of $\Omega_{\Delta n}$. The estimated parameters for
all the PDFs are listed in Table \ref{Tb:FittingPDF:Qn}.
Fig.~\ref{Fig:Qn}(b) demonstrates the fits for four typical PDFs
with $\Delta n=2, 8, 32, 128$. The $q$-Gamma fits the data with good
quality for small values of $\Delta{n}$. For larger values of
$\Delta{n}$, the fits deviate considerably from the empirical data.
Again, the Cram\'{e}r-von-Mises tests is also used for judging the
goodness-of-fit. The results are favorable only when
$\Delta{n}\leqslant8$ at the significance level 0.01. The value of
$C_{M}^{2}$ increases with the increase of $\Delta{n}$ when
$\Delta{n}$ is larger than 8. So we can declare that the
applicability of the $q$-Gamma functions for multiple trades is
restricted to the transaction number $\Delta{n}\leqslant8$.

\begin{table}[htp]
 \caption{\label{Tb:FittingPDF:Qn} Calibration of the $q$-Gamma model to trading volumes at different event timescales.}
 \medskip
 \centering
 \begin{tabular}{ccccccc}
 \hline \hline
  $\Delta{n}$ && $\theta$ & $\beta$ &  $q$  & $\chi$  &  $\alpha^{'}$   \\%
    \hline
    1         &&    0.07  &  1.52   &  1.22 &  0.57   &  2.03           \\%
    2         &&    0.04  &  2.68   &  1.17 &  0.47   &  2.20           \\%
    4         &&    0.07  &  2.56   &  1.16 &  0.26   &  2.60           \\%
    8         &&    0.07  &  3.28   &  1.14 &  0.29   &  2.86           \\%
    16        &&    0.09  &  3.45   &  1.12 &  0.45   &  3.88           \\%
    32        &&    0.09  &  3.37   &  1.12 &  0.65   &  3.96           \\%
    64        &&    0.12  &  3.41   &  1.10 &  0.87   &  5.59           \\%
    128       &&    0.11  &  3.39   &  1.10 &  0.92   &  5.61           \\%
    256       &&    0.15  &  3.52   &  1.07 &  1.04   &  9.77           \\%
    \hline \hline
 \end{tabular}
\end{table}

In order to determine the tail exponents, the six estimators
mentioned in the previous sections are utilized. The resultant tail
exponents are presented in Table \ref{Tb:FittingPDF:Qn_tail}. The
$\alpha$ values obtained from LSE, CSNE, FAE and RKE are greater
than two, while those obtained from HE and MSE are less than two for
small $\Delta{n}$. There is an overall tendency that $\alpha$
obtained from different method increases with $\Delta{n}$. This can
be considered as a significance of the central limit theorem meaning
that for $\Delta{n} \to \infty$, the distribution converges to a
Gaussian. The fact that the limit distribution is not a L{\'e}vy
stable one is another indication that $\alpha > 2$.

\begin{table*}[htp]
 \caption{\label{Tb:FittingPDF:Qn_tail} Estimating tail exponents of trading volumes at different event timescales.}
 \medskip
 \centering
 \begin{tabular}{cccccccccc}
 \hline \hline
  $\Delta{n}$&&   $\alpha^{'}$ &&  HE            &   MSE            &  CSNE         &  LSE           & FAE          &    RKE         \\%
    \hline
    1        &&      2.03      &&  $1.54\pm0.07$ &  $1.75\pm0.003$  &  $2.23\pm0.23$ & $2.33\pm0.08$ & $2.59\pm0.39$& $2.13\pm0.23$   \\%
    2        &&      2.20      &&  $1.75\pm0.02$ &  $1.81\pm0.006$  &  $2.24\pm0.34$ & $2.54\pm0.09$ & $2.95\pm0.22$& $2.31\pm0.20$   \\%
    4        &&      2.60      &&  $2.07\pm0.02$ &  $1.87\pm0.005$  &  $2.21\pm0.07$ & $2.52\pm0.05$ & $3.10\pm0.52$& $2.35\pm0.17$   \\%
    8        &&      2.86      &&  $2.32\pm0.03$ &  $1.94\pm0.005$  &  $2.63\pm0.16$ & $2.81\pm0.07$ & $3.09\pm0.68$& $2.80\pm0.19$   \\%
    16       &&      3.88      &&  $2.49\pm0.04$ &  $2.01\pm0.010$  &  $2.80\pm0.13$ & $2.87\pm0.07$ & $3.03\pm0.62$& $3.20\pm0.34$   \\%
    32       &&      3.96      &&  $2.93\pm0.07$ &  $2.09\pm0.018$  &  $2.97\pm0.20$ & $2.92\pm0.08$ & $3.12\pm0.46$& $3.57\pm0.44$   \\%
    64       &&      5.59      &&  $3.36\pm0.15$ &  $2.18\pm0.011$  &  $3.33\pm0.22$ & $3.18\pm0.10$ & $3.30\pm0.45$& $3.97\pm0.58$   \\%
    128      &&      5.61      &&  $3.97\pm0.43$ &  $2.27\pm0.026$  &  $3.84\pm0.23$ & $3.26\pm0.13$ & $3.65\pm0.57$& $4.20\pm0.69$   \\%
    256      &&      9.77      &&  $4.37\pm0.64$ &  $2.39\pm0.042$  &  $4.51\pm0.47$ & $3.69\pm0.16$ & $4.05\pm0.68$& $5.19\pm0.77$   \\%
    \hline \hline
 \end{tabular}
\end{table*}

\section{Conclusion}
\label{S1:Conclusion}

The distributions of trade sizes and trading volumes are
investigated based on the limit order book data of 22 liquid Chinese
stocks listed on the Shenzhen Stock Exchange in the whole year 2003.
The size distribution of trades for individual stocks exhibits
jumps, which is caused by the fact that traders prefer to place
orders with the size being certain numbers. The empirical PDFs of
trading volumes at different timescales $\Delta{t}$ ranging from 1
min to 240 min can be modeled by the $q$-Gamma functions. In
contrast, the empirical PDFs of trading volumes for multiple trades
can be fitted by the $q$-Gamma functions only for small numbers of
trades, $\Delta{n}\leqslant8$. All the empirical PDFs exhibit
power-law tails. In order to determine the tail exponents, we
adopted six estimators (HE, MSE, LSE, CSNE, FAE and RKE). The
estimated tail exponents using LSE, CSNE, FAE and RKE are greater
than two, while those obtained from HE and MSE are less than two
when $\Delta{t}$ or $\Delta{n}$ is small. Since HE and MSE may
underestimate the tail exponents, we conclude that the tail
exponents of trade sizes and trading volumes of Chinese stocks are
well outside the L{\'{e}}vy regime.

\begin{acknowledgement}

This work was partly supported by the National Natural Science
Foundation of China (Nos. 70501011 and 70502007), the Fok Ying Tong
Education Foundation (No. 101086), the Program for New Century
Excellent Talents in University (No. NCET-07-0288), and the China
Scholarship Council (No. 2008674017).
\end{acknowledgement}

\bibliographystyle{epj}
\bibliography{E:/Papers/Auxiliary/Bibliography}

\end{document}